\documentclass[twocolumn, twocolappendix]{aastex631}

\usepackage{xspace}
\usepackage{color, soul}
\usepackage{multirow}
\usepackage{amsmath}
\usepackage{textcomp}
\usepackage{orcidlink}
\usepackage[utf8]{inputenc}

\submitjournal{ApJ}
\received{December 30, 2025}
\accepted{March 17, 2026}

\shortauthors{Zhu et al.}
\shorttitle{Diffuse X-ray Emission in the Sagittarius C Complex}

\begin{document}

\title{Diffuse X-ray Emission in the Sagittarius C Complex}

\correspondingauthor{Zhenlin Zhu}
\email{z.zhu@astro.ucla.edu}

\author{Zhenlin Zhu\,\orcidlink{0000-0001-8812-8284}}
\affiliation{Department of Physics and Astronomy, University of California, Los Angeles, CA, 90095-1547, USA}

\author{Mark R. Morris\,\orcidlink{0000-0002-6753-2066}}
\affiliation{Department of Physics and Astronomy, University of California, Los Angeles, CA, 90095-1547, USA}

\author{Gabriele Ponti\,\orcidlink{0000-0003-0293-3608}}
\affiliation{INAF – Osservatorio Astronomico di Brera, Via E. Bianchi 46, 23807 Merate, Italy}
\affiliation{Max-Planck-Institut für Extraterrestriche Physik, Gießenbachstraße 1, 85748 Garching, Germany}
\affiliation{Como Lake Center for Astrophysics (CLAP), DiSAT, Università degli Studi dell’Insubria, via Valleggio 11, I-22100 Como, Italy}

\author{Ping Zhou\,\orcidlink{0000-0002-5683-822X}}
\affiliation{School of Astronomy and Space Science, Nanjing University, Nanjing 210023, People’s Republic of China}

\begin{abstract}

The Sagittarius C (Sgr C) complex, located on the western edge of the Central Molecular Zone (CMZ), hosts a mixture of star-forming and non-thermal activity whose X-ray properties remain poorly understood. Using deep archival {\it Chandra} and {\it XMM-Newton} observations, we resolve the diffuse X-ray emission in Sgr C into two components: an H\,\textsc{ii} region coincident with the radio peak and a brighter diffuse feature located to its southwest. Spatially resolved spectroscopy reveals the presence of a soft (kT $\leq$ 1 keV) plasma with metal abundances consistent with the elevated metallicity expected in the CMZ in both regions, along with a harder ($\sim$8 keV) thermal component within the H,\textsc{ii} region. The observed diffuse X-ray emission and its association with an expanding [C\,\textsc{ii}] shell suggest that the hot gas may originate from a young supernova remnant (SNR) embedded in the H\,\textsc{ii} region. 
Under this interpretation, the inferred shock velocity ($\sim 800~\mathrm{km~s^{-1}}$) and SNR age ($\geq 1.7~\mathrm{kyr}$) are consistent with a core-collapse SNR in the Galactic Center. 
These results reveal Sgr C as a potential host of a SNR and highlight the complex interplay between massive-star feedback, magnetic fields, and molecular gas in the CMZ.

\end{abstract}


\section{Introduction} 

The inner regions of our Galaxy exhibit prominent diffuse X-ray emission, arising from a combination of truly diffuse hot plasma and unresolved point sources. 
The Galactic Center diffuse emission has been observed extensively with {\it Chandra}, {\it XMM-Newton}, and Suzaku, revealing a complex mixture of thermal and non-thermal components associated with supernova activity, stellar winds, and possible past outbursts of Sgr A* \citep{Koyama1996, Muno2004, Ponti2015}.
A particularly striking feature of the Galactic Center diffuse emission is the presence of strong 6.4 keV Fe K$\alpha$ line emission, observed from numerous dense molecular clouds in the central molecular zone \citep[CMZ;][]{Morris1996}, such as Sgr B2, Sgr C, and the “Bridge” region \citep[See][for a review]{Ponti2013}. 
This neutral Fe line is widely interpreted as reflection or fluorescent emission, produced when hard X-ray photons and cosmic rays — most likely originating from past outbursts of Sgr A*—irradiate cold molecular gas, leading to photoionization and subsequent K$\alpha$ fluorescence \citep{Koyama1996, Murakami2000, Sunyaev1998, Terrier2010, Clavel2013, Anastasopoulou2025}. 
Temporal variations in the 6.4 keV line flux measured over the past two decades and the polarization of the associated X-ray continuum provide strong evidence for this “light-echo” scenario, thereby offering a unique probe of the recent activity history of our Galaxy’s central supermassive black hole.
On larger scales along the Galactic plane, the Galactic Ridge X-ray Emission (GRXE), characteristic of a $10^{8}$ K optically thin thermal plasma, has been traced by the $\sim$6.7 keV He-like iron (Fe XXV) line \citep{Worrall1982, Koyama1986, Yamauchi1993}. 
The GRXE displays a strong correlation with the near-infrared stellar distribution, suggesting that it is largely due to the integrated contribution of faint accreting binaries and coronally active stars \citep{Revnivtsev2006, Revnivtsev2009, Warwick2014, Anastasopoulou2023}.
To fully characterize the diffuse X-ray emission in the Galactic Center, it is essential to account for both the reflected component from molecular clouds and the large-scale GRXE.

\begin{deluxetable*}{lllcccl}[!htbp]
\tablecaption{X-Ray Observational Details\label{tab:xray_obs}}
\tablewidth{0pt}
\tablehead{
\colhead{Telescope} & \colhead{ObsID} & \colhead{Instrument} & 
\colhead{R.A.} & \colhead{Decl.} & 
\colhead{Start Date} & \colhead{Valid Exp.} \\
 & & & (deg) & (deg) & & (ks)
}
\startdata
XMM-Newton
& 112970701
 & EPIC-MOS1 & 266.20033 & -29.36392 & 2000-09-11 & 23.0 \\
  & & EPIC-MOS2 &  &  &  & 22.9 \\
  & & EPIC-PN   &  &  &  & 16.8 \\
 & 112970801
 & EPIC-MOS1 & 266.03325 & -29.59864 & 2000-09-21 & 23.2 \\ 
  & & EPIC-MOS2 &  &  &  & 23.1 \\
  & & EPIC-PN   &  &  &  & 11.3 \\
 & 302883101
 & EPIC-MOS1 & 266.14896 & -29.3345 & 2006-02-27 & 11.0 \\
  & & EPIC-MOS2 &  &  &  & 11.3 \\
  & & EPIC-PN   &  &  &  & 7.0 \\
 & 302884501
 & EPIC-MOS1 & 266.1855 & -29.38267 & 2006-09-09 & 8.1 \\
  & & EPIC-MOS2 &  &  &  & 8.2 \\
  & & EPIC-PN   &  &  &  & 6.0 \\
 & 504940701
 & EPIC-MOS1 & 266.18608 & -29.38256 & 2007-09-06 & 6.4 \\
  & & EPIC-MOS2 &  &  &  & 6.5 \\
  & & EPIC-PN   &  &  &  & 3.9 \\
 & 511001301
 & EPIC-MOS1 & 266.14913 & -29.33486 & 2008-03-04 & 5.3 \\
  & & EPIC-MOS2 &  &  &  & 5.1 \\
  & & EPIC-PN   &  &  &  & 2.9 \\
 & 511001401
 & EPIC-MOS1 & 266.18462 & -29.38342 & 2008-09-27 & 6.3 \\
  & & EPIC-MOS2 &  &  &  & 6.5 \\
  & & EPIC-PN   &  &  &  & 4.4 \\
 & 694640201
 & EPIC-MOS1 & 266.28388 & -29.31928 & 2012-08-30 & 45.7 \\
  & & EPIC-MOS2 &  &  &  & 45.9 \\
  & & EPIC-PN   &  &  &  & 40.0 \\
 & 694640101
 & EPIC-MOS1 & 266.13308 & -29.53292 & 2012-09-07 & 41.2 \\
  & & EPIC-MOS2 &  &  &  & 41.5 \\
  & & EPIC-PN   &  &  &  & 34.6 \\
 & 694640901
 & EPIC-MOS1 & 266.2115 & -29.42175 & 2012-09-12 & 42.7 \\
  & & EPIC-MOS2 &  &  &  & 43.6 \\
  & & EPIC-PN   &  &  &  & 37.4 \\
 & 862470101
 & EPIC-MOS1 & 266.13758 & -29.52967 & 2020-09-06 & 48.2 \\
  & & EPIC-MOS2 &  &  &  & 48.4 \\
  & & EPIC-PN   &  &  &  & 39.8 \\
 & 862471301
 & EPIC-MOS1 & 266.28317 & -29.31881 & 2020-09-11 & 7.8 \\
  & & EPIC-MOS2 &  &  &  & 7.8 \\
  & & EPIC-PN   &  &  &  & 5.2 \\
 & 862470301
 & EPIC-MOS1 & 266.28317 & -29.31886 & 2020-09-11 & 44.0 \\
  & & EPIC-MOS2 &  &  &  & 44.0 \\
  & & EPIC-PN   &  &  &  & 36.2 \\
 & 862470201
 & EPIC-MOS1 & 266.20992 & -29.42244 & 2020-10-04 & 39.8 \\
  & & EPIC-MOS2 &  &  &  & 40.5 \\
  & & EPIC-PN   &  &  &  & 31.3 \\
\hline
Chandra 
&  2272 & ACIS-I & 266.05217 & -29.43981 & 2001-07-20 & 10.5\\ 
&  5892 & ACIS-I & 266.08458 & -29.43504 & 2005-07-22 & 95.3 \\
&  16174 & ACIS-I & 266.09643 & -29.40234 & 2014-07-29 & 28.7 \\
&  16642 & ACIS-I & 266.09542 & -29.40099 & 2014-08-01 & 28.4 \\
&  16643 & ACIS-I & 266.09570 & -29.40127 & 2014-08-03 & 34.0 
\enddata
\end{deluxetable*}

\begin{figure*}[!tbh]
\centering
\includegraphics[width=0.45\textwidth]{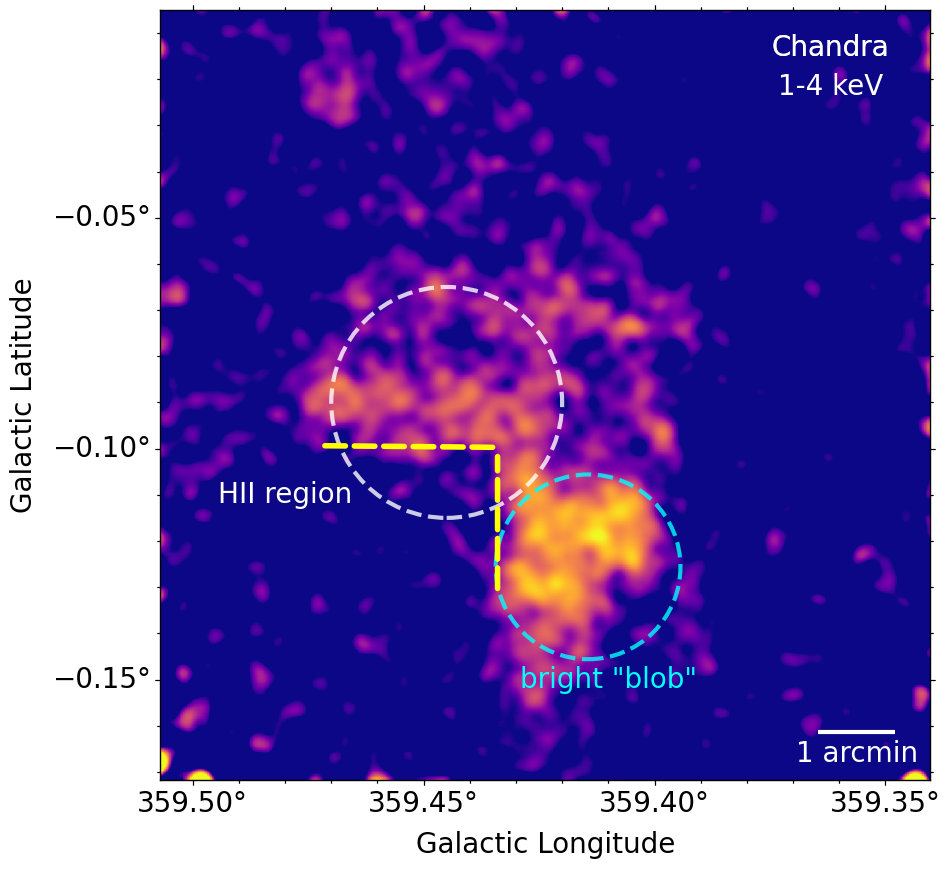}
\includegraphics[width=0.45\textwidth]{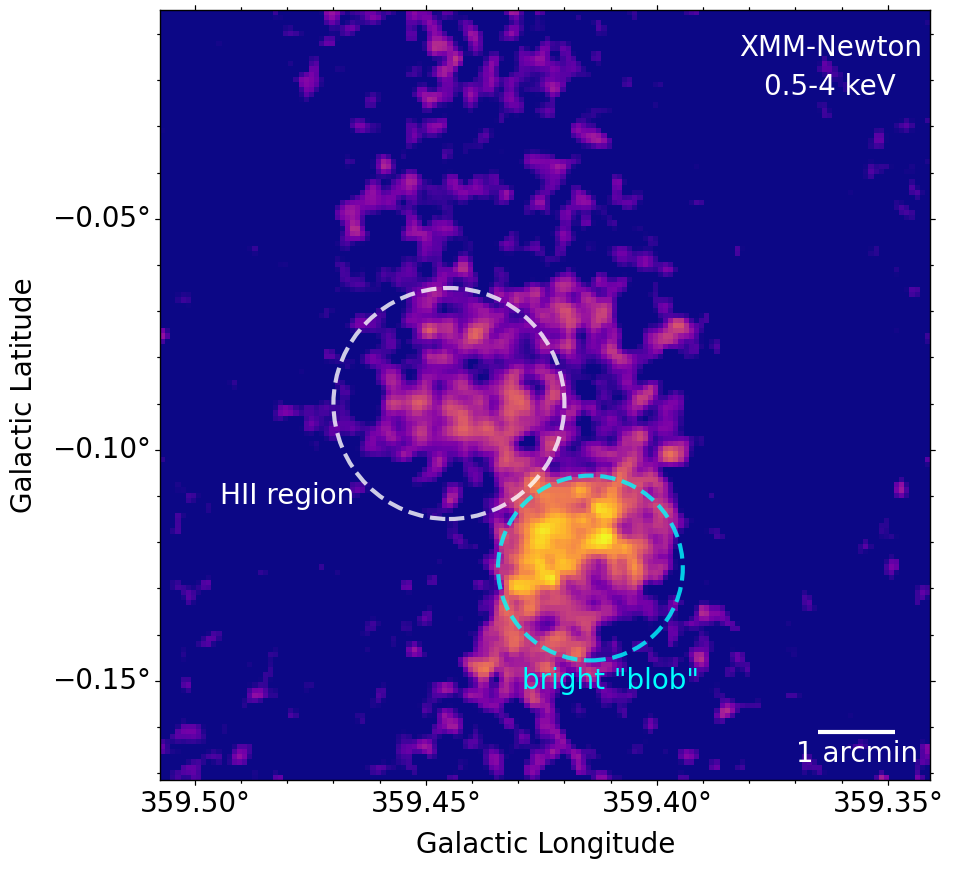}
\caption{Exposure- and vignetting-corrected X-ray flux images of the Sgr C region. {\it Left}: {\it Chandra} flux image in the 1--4 keV band. For visualization purpose, the image has been smoothed with a gaussian kernel of 11 $\times$ 11 pixels (5.5\arcsec x 5.5 \arcsec). {\it Right}: {\it XMM-Newton} flux image in the 0.5-4 keV band. The dashed white circle in both images shows the location and approximate periphery of the Sgr C H\,\textsc{ii} region, while the cyan circles show the newly-discovered feature -- the bright ``Blob". In the {\it Chandra} image, we mark the edges of the X-ray emission with yellow dashed lines, which show an excellent spatial coincidence with the edge of the adjacent distribution of foreground cold dust, as shown in Figure~\ref{fig:infrared}. }
\label{fig:xray_fig}
\end{figure*}

Sagittarius~C (hereafter Sgr~C) is a massive and active star-forming region located at the western side of the Central Molecular Zone (CMZ), and it thus serves as a useful laboratory for testing theories of star formation in the relatively extreme environment of the CMZ \citep{Tsuboi1999, Kendrew2013, Henshaw+23}. 
However, in contrast to Sagittarius~B2 (Sgr~B2), which lies on the opposite side of the CMZ and has been extensively studied, Sgr~C has received relatively little attention until recent years. 
The 1.28~GHz MeerKAT Galactic Center Mosaic \citep{Heywood2022} provides a detailed radio view of Sgr~C, revealing intense emission at the location of the H\,\textsc{ii} region and a spectacular non-thermal filament (NTF) extending toward the north\footnote{All cardinal directions in this paper are defined in the Galactic coordinate system.}. 
In 2023, observations from the {\it James Webb Space Telescope} (JWST) mapped the Sgr~C complex, revealing bright hydrogen recombination line Br-$\alpha$ emission filling the H\,\textsc{ii} region \citep{Crowe2025, Bally2025}. 
More recently, \citet{Riquelme2025} reported an expanding shell feature surrounding the Sgr C H\,\textsc{ii} region in the [C\,\textsc{ii}] emission map obtained with the {\it GREAT} instrument on {\it SOFIA}. 
Those authors argue that their estimated expansion velocity, 23.4 km s$^{-1}$, cannot be explained solely by stellar feedback within the H\,\textsc{ii} region, and raise the possibility of a recent supernova within the H\,\textsc{ii} region. 
Furthermore, using 214~$\mu$m polarimetric observations from {\it SOFIA}/HAWC+, \citet{Zhao2025} traced the magnetic field structure and found its orientation to be tangential to the surface of the Sgr~C H\,\textsc{ii} region.

The first X-ray observations of the Sgr~C region were made with {\it Suzaku} by \citet{Tsuru2009}. who presented images and X-ray spectra of both the central diffuse emission and an extended north-south feature they referred to as the ``Chimney'', but which we will hereafter refer to as the Sgr C ``Channel''\footnote{We suggest this name change here to avoid confusion with the much larger-scale Galactic center Chimneys discovered more recently in X-rays and radio \citep{Ponti2019,Ponti2021,Heywood+19}. The possible relationship of the Channel to the chimney is discussed below in Section \ref{sec:channel}. }. 
They estimated the total thermal energy of the Sgr~C complex to be $1.4\times10^{50}$~erg, and were the first to propose the presence of a supernova remnant (SNR) within this region. 
However, due to the limited angular resolution of {\it Suzaku} and to the offset of the peak of the diffuse X-ray emission from the H\,\textsc{ii} region, an association of the emitting plasma with the H\,\textsc{ii} region could not be firmly established.

More recently, \citet{Wang2021} presented a {\it Chandra} survey of the central $2^\circ \times 4^\circ$ field of the Galaxy, revealing a much clearer morphology of the diffuse X-ray emission in Sgr~C. 
In addition, \citet{Chuard2018} analyzed {\it Chandra} and {\it XMM-Newton} data of Sgr~C, mapping the fluorescent Fe~K$\alpha$ emission and presenting the {\it XMM-Newton} spectrum based on an effective exposure of $\sim$130~ks. 
Their analysis, however, primarily focused on tracing the past high-energy activity of Sgr~A$^{*}$ rather than of the Sgr~C region itself.

{\it Chandra} and {\it XMM-Newton} have been actively monitoring the Galactic Center since 1999. 
The archival {\it Chandra} observations cover the central $4^\circ \times 6^\circ$ region of the Galaxy, providing high-resolution imaging of the diffuse X-ray emission. 
Meanwhile, the ongoing {\it XMM-Newton} Heritage Survey (PI: G. Ponti) is mapping a much wider field of view ($350^\circ < l < 7^\circ,\;-1^\circ < b < 1^\circ$) across the inner Galactic disk, with the primary goal of tracing the flow of hot baryons that drive large-scale energetic phenomena such as the Galactic center chimneys \citep{Ponti2019, Ponti2021}, the {\it eROSITA} bubbles \citep{Predehl2020}, and the Fermi bubbles \citep{Su2010}. 
By using both {\it Chandra} and {\it XMM-Newton} observations, we obtain the deepest exposure to date for Sgr C, enabling high-quality, spatially resolved spectroscopy of this region.

In this work, we make use of archival {\it Chandra} and {\it XMM-Newton} observations to re-visit the environment of the Sgr C H\,\textsc{ii} region and molecular cloud. 
Section~\ref{sec:obs} describes the observations employed in this study and the details of the data reduction procedures. 
In Section~\ref{sec:image}, we present the X-ray images across different energy bands along with corresponding hardness ratio maps, then in Section \ref{sec:spectra}, we show the extracted spectra and their best-fit models.
Finally, we discuss and interpret our results in Section \ref{sec:discussion} and summarize our conclusions in Section \ref{sec:conclusion}.
Throughout this paper, we adopt an average distance of 8.1 kpc to the Galactic Center \citep{Do2019,GRAVITY2022}, for which 1\arcmin ~corresponds to 2.36 pc.

\section{Observations and data reduction} \label{sec:obs}
\subsection{Chandra}
The Galactic Center has been frequently observed by {\it Chandra}'s Advanced CCD Imaging Spectrometer (ACIS) since 1999.
To map the Sgr C region, we utilized 5 observations taken in ACIS-I VFAINT mode, which accumulated a total of 197 ks of clean exposure (See Table \ref{tab:xray_obs}).

All data were reprocessed using the CIAO data analysis package, version 4.16, and the latest calibration database (CALDB 4.11.5) distributed by the Chandra X-ray Observatory Center.
We removed the cosmic rays and bad pixels from all level-1 event files using the CIAO tool \texttt{chandra\_repro} with VFAINT mode background event filtering.
For each observation, we extracted the lightcurve in the 9--12 keV band to detect possible contamination from flare events.
To remove the time intervals with anomalous background, we applied the CIAO tool \texttt{deflare} to filter out the times where the background rates exceed $\pm 2 \sigma$ of the average value.

We utilized stowed background files to estimate the non-X-ray background (NXB). 
All stowed background event files were combined and reprocessed with \texttt{acis\_process\_events} using the latest gain calibration files.
For each observation, we scaled the NXB to match the 9-12 keV count rate of the observation.

We extracted and combined the 1--4 keV count images using \texttt{merge\_obs}. 
Background images were generated from the stowed observation files.
We calculated the exposure maps using a weighted spectrum file generated by \texttt{make\_instmap\_weighted}, where the spectral model is an absorbed \texttt{apec} model with kT = 1 keV, which is the average plasma temperature measured by the previous {\it Suzaku} study \citep{Tsuru2009}.
Point sources were identified using the CIAO tool \texttt{wavdetect}, masked from the images, and the masked regions were refilled using \texttt{dmfilth}.
After correcting the NXB-subtracted count image with the combined exposure map and smoothing it with a Gaussian kernel of 11 × 11 pixels (5.5 \arcsec x 5.5 \arcsec), we obtained the \textit{Chandra} flux map shown in the left panel of Figure~\ref{fig:xray_fig}.

\begin{figure}
\centering
\includegraphics[width=0.45\textwidth]{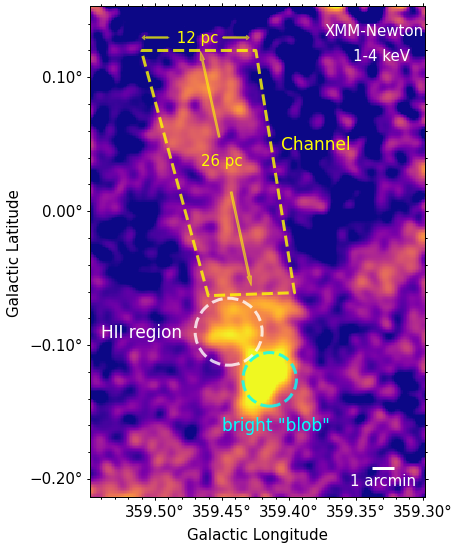}
\caption{A detailed {\it XMM-Newton} view of the “Channel,” a diffuse X-ray structure extending $\sim$26 pc northward.}
\label{fig:chimney}
\end{figure}

\begin{figure*}[!tbh]
\centering
\includegraphics[width=0.47\textwidth]{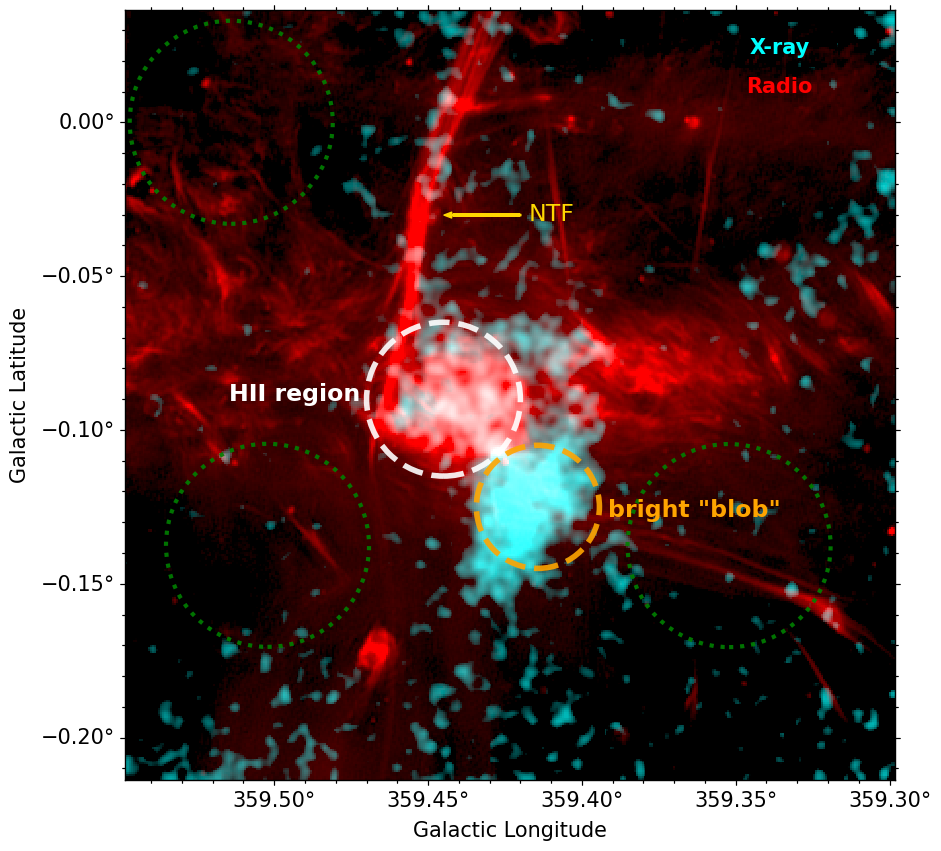}
\includegraphics[width=0.45\textwidth]{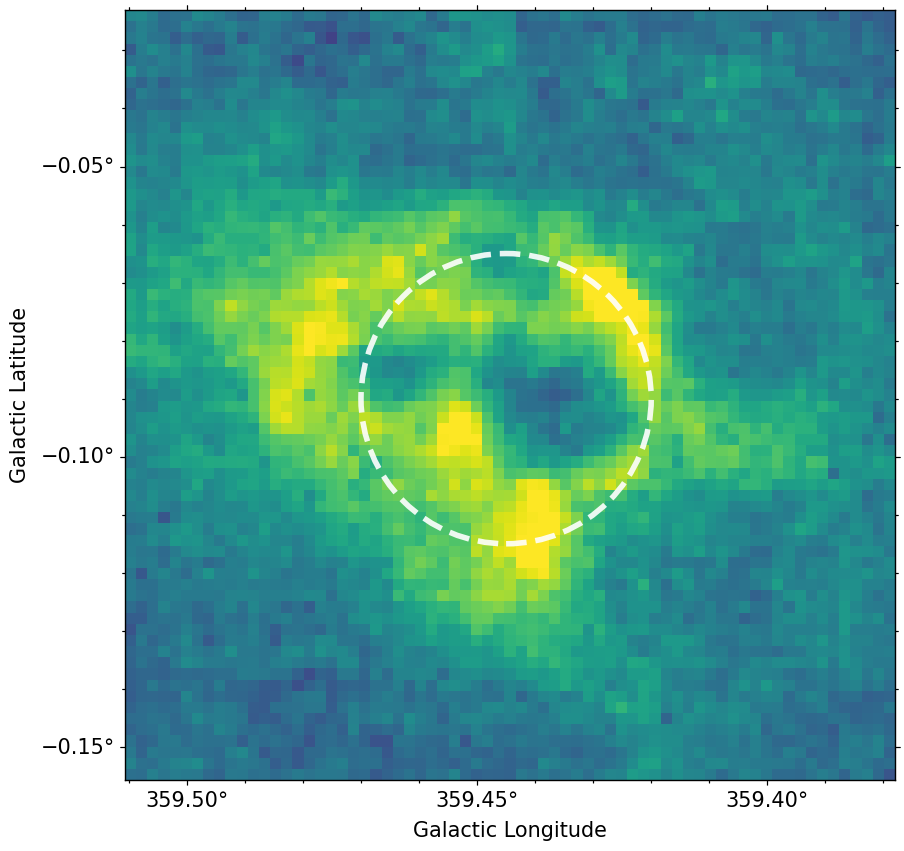}
\caption{Multi-wavelength view of Sgr C. {\it Left:} Radio and X-ray overlay of Sgr C. The radio emission of MeerKAT is shown in red, while the cyan shows the 0.5--2 keV {\it XMM-Newton} X-ray emission. The 3 dotted green circles denote the selected local background regions. The white dashed circle encloses the H\,\textsc{ii} region, while the orange dashed circle marks the Blob region. {\it Right:} the CMZ in [C\,\textsc{ii}] Survey image of the $-41$ km~s$^{-1}$ velocity channel observed using SOFIA/upGREAT \citep{Riquelme2025}. The H \textsc{ii} region is marked by a white dashed circle.} 
\label{fig:multi_fig}
\end{figure*}

\subsection{XMM-Newton}
The Sgr C region was observed for 120\,ks by the {\it XMM-Newton} European Photon Imaging Camera (EPIC) MOS+pn from 2000 to 2020 (see Table~\ref{tab:xray_obs}). We reduced the data using the XMM-Newton Science Analysis System (SAS) v21.0.0. MOS and pn event files were obtained from the observation data files with the tasks \texttt{emchain} and \texttt{epchain}. The out-of-time event file of pn was also produced by \texttt{epchain}.

The data were cleaned for periods of high background using the tasks \texttt{mos-filter} and \texttt{pn-filter}. The more recent observations taken in 2020 suffer from heavy solar flares, especially for EPIC-pn observations. After this filtering, the net exposure times of detectors MOS1, MOS2, and pn were $\sim$332\,ks, $\sim$332\,ks, and 261\,ks, respectively. Following standard filtering techniques, we kept the single, double, triple, and quadruple events in the MOS data (pattern $\leq$ 12) and restricted the pn data to single and double events only (pattern $\leq4$). Additionally, we also excluded all the CCDs in the anomalous state \citep[for more details, see][]{Kuntz2008}.

To produce the 0.5--4\,keV {\it XMM-Newton} flux image, we created the count maps from the cleaned event files and generated the vignetting-corrected exposure maps with \texttt{eexpmap}. 
Point-like sources were detected using the task \texttt{edetect\_chain}, and all were excluded from subsequent analyzes.
We further rescaled the Filter Wheel Closed (FWC) data to match the 9–12 keV count rate of our observations, and used them as instrumental background event files.

The background-subtracted and vignetting-corrected {\it XMM-Newton} image in the 0.5–4.0 keV band shown in Figure~\ref{fig:xray_fig} was produced with a binning of 4 physical pixels, yielding an effective angular resolution of 
4\arcsec.
For visualization purposes, we refilled the point-source regions using the CIAO task \texttt{dmfilth}.

\section{X-ray imaging} 
\label{sec:image}
\subsection{Morphology and properties}
\label{sec:morphology}
As shown in Figure~\ref{fig:xray_fig}, the entire structure exhibits an inverted L-shaped morphology. 
Prominent emission is spatially associated with the Sgr~C H\,\textsc{ii} region; however, the brightest component is located toward the southwest of the H\,\textsc{ii} region, offset by approximately 2.7\arcmin. 
We refer to this diffuse feature as the bright ``Blob,'' enclosed within a roughly circular region of radius 1.2\arcmin. 

The H\,\textsc{ii} region appears as a separate X-ray source largely coincident with the radio H\,\textsc{ii} region, but the projected image raises the possibility that it is connected to the Blob.  
The H\,\textsc{ii} region counterpart exhibits a surface brightness of $7.7\times 10^{-13} ~$erg~cm$^{-2}~$s$^{-1}$~arcmin$^{-2}$ in 1.5-8 keV, whereas the Blob is approximately $\sim$70\% brighter, with $1.3\times 10^{-12} ~$erg~cm$^{-2}$~s$^{-1}$~arcmin$^{-2}$.
An X-ray dark region is observed south of the H,\textsc{ii} region, bounded by two well-defined edges (yellow dashed lines in Figure \ref{fig:xray_fig}), likely caused by strong absorption from foreground cold dust (see Figure~\ref{fig:infrared} and Section~\ref{sec:diss-blob}).

In addition, as shown in Figure \ref{fig:chimney}, diffuse emission constituting the Channel extends northward from the H\,\textsc{ii} region, as was first reported by \citet{Tsuru2009} (and referred to by them as the ``chimney''). 
Taking advantage of the larger field of view covered by {\it XMM-Newton}, we measure this feature to extend to a length of $\sim$26 pc, although its surface brightness is significantly weaker than that of both the H\,\textsc{ii} region and the bright Blob.
As shown in Figure~\ref{fig:multi_fig}, the prominent radio non-thermal filament \citep[NTF; see][for details]{Liszt1985,Liszt1995} appears to overlap spatially with the Channel although it exhibits a slightly different orientation. 
Without a much higher resolution and higher signal-to-noise X-ray image, it remains uncertain whether these two features are physically connected or are projected as a chance alignment. The properties of the Channel are further discussed in Section \ref{sec:channel}.

\subsection{Hardness Ratio}
\label{sec:hr}

\begin{figure*}[!tbh]
\centering
\includegraphics[width=0.85\textwidth]{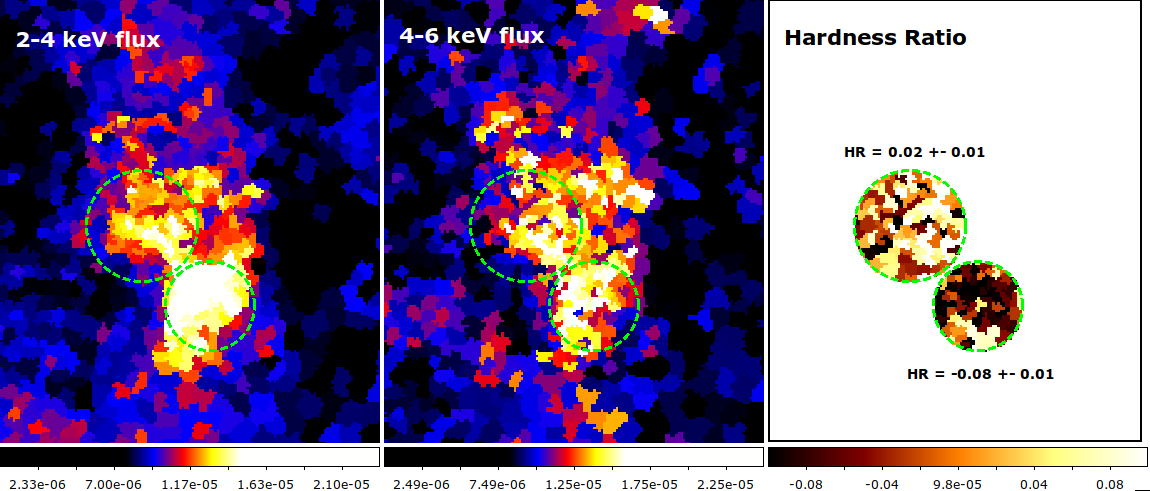}
\caption{Adaptively-binned {\it XMM-Newton} flux maps and hardness ratio map. 
The dashed circles mark the locations of the H\,\textsc{ii} region and the Blob. 
The signal-to-noise ratio for each bin is set to be higher than 10. }
\label{fig:bin_maps}
\end{figure*}

For {\it XMM-Newton}, a contour binning algorithm \citep{Sanders2006} was applied to this flux map, to create a binned image with each bin reaching a signal-to-noise ratio (S/N) of 10 (see Figure \ref{fig:bin_maps}). 

The hardness ratio (HR) is defined as $HR = (F_{H} - F_{L}) / (F_{H} + F_{L})$, where $F_{H}$ and $F_{L}$ refer to the photon fluxes in the high energy (4-6 keV) and low energy (2-4 keV) bands, respectively.

Using the adaptively binned flux maps, we constructed a hardness-ratio map with the same binning scheme, also shown in Figure \ref{fig:bin_maps}. 
The mean HRs of the H\,\textsc{ii} region and the bright Blob are 0.02$\pm$0.01 and $-0.08\pm0.01$, respectively, revealing that the  H\,\textsc{ii} region emission is significantly harder than that of the Blob.

\begin{figure*}[!tbh]
\centering
\includegraphics[width=0.48\textwidth]{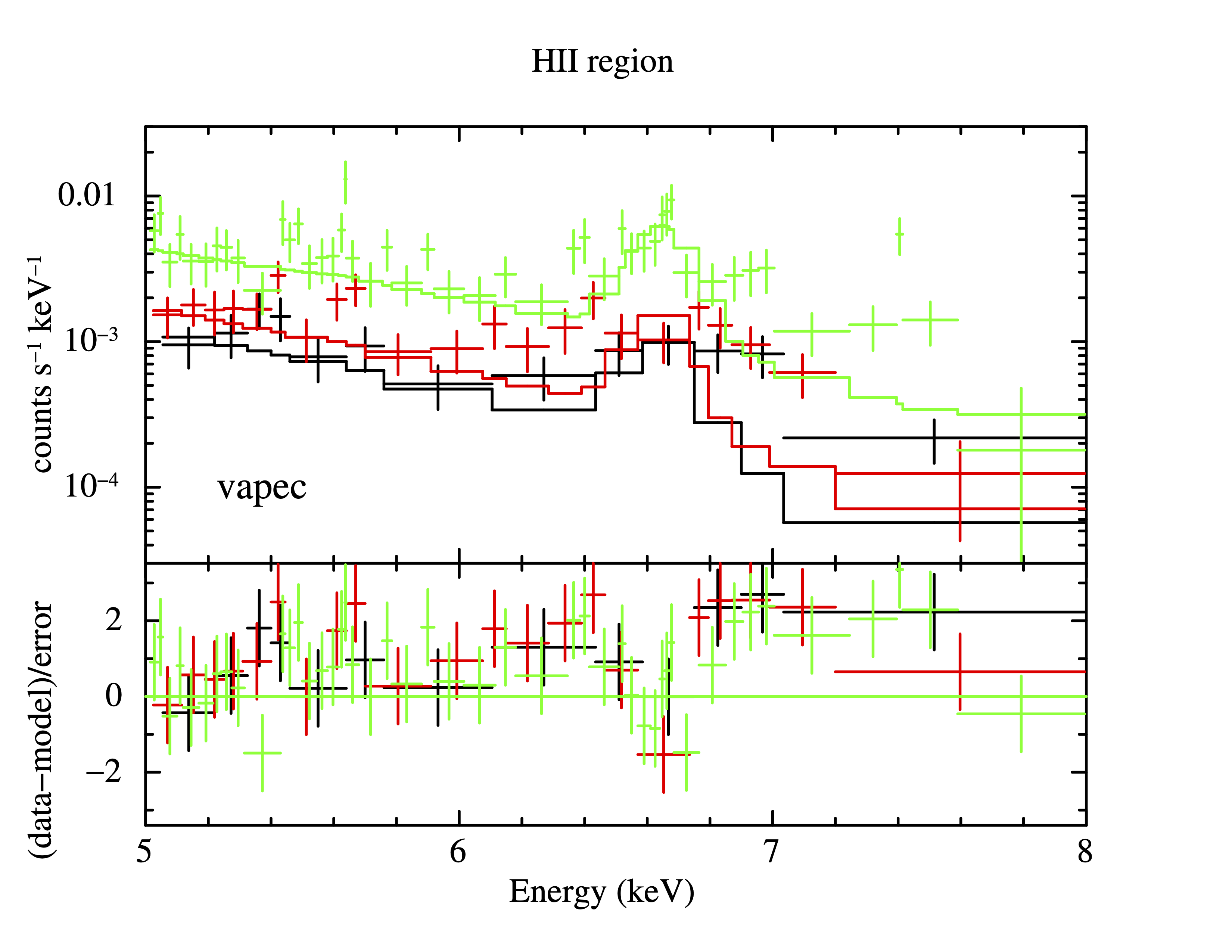}
\includegraphics[width=0.48\textwidth]{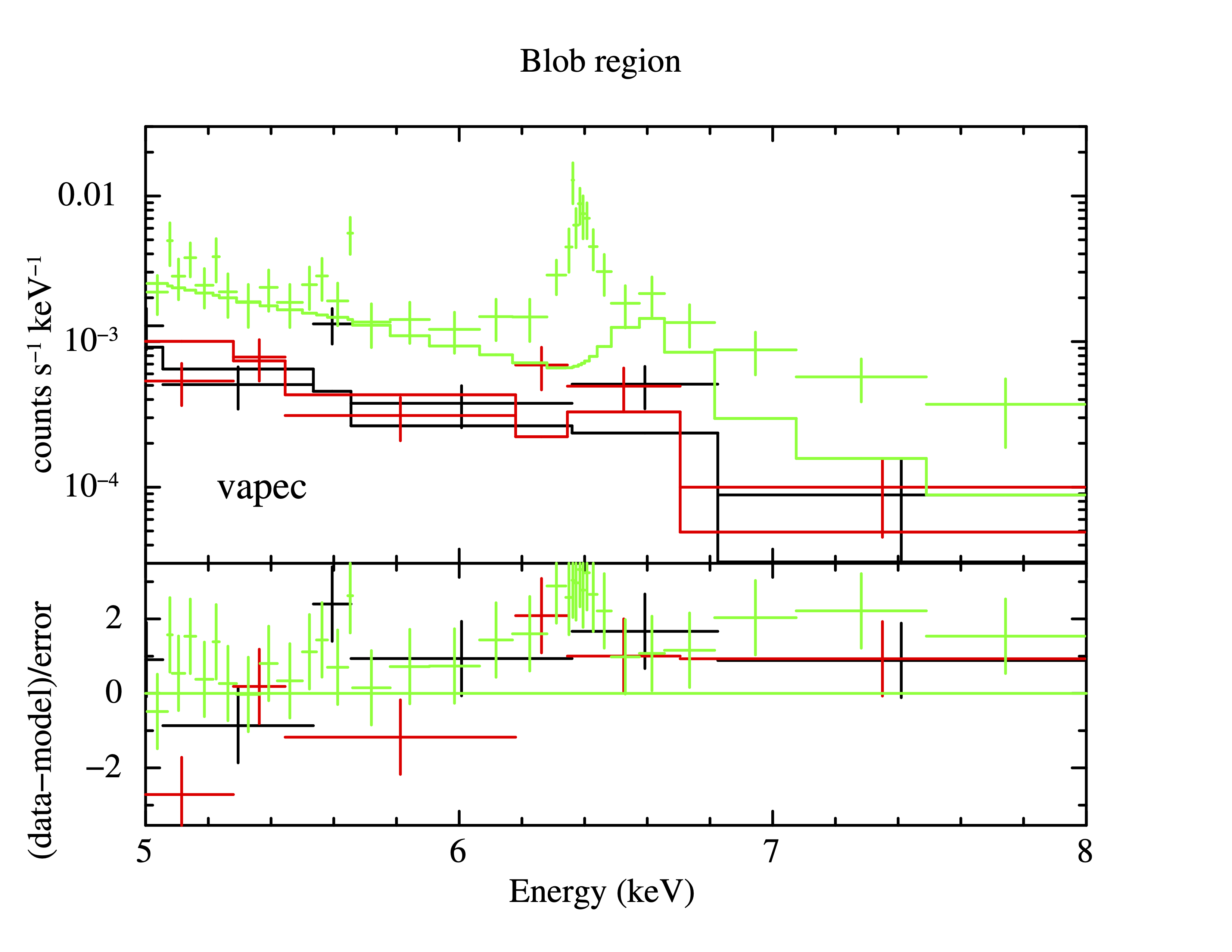}
\includegraphics[width=0.48\textwidth]{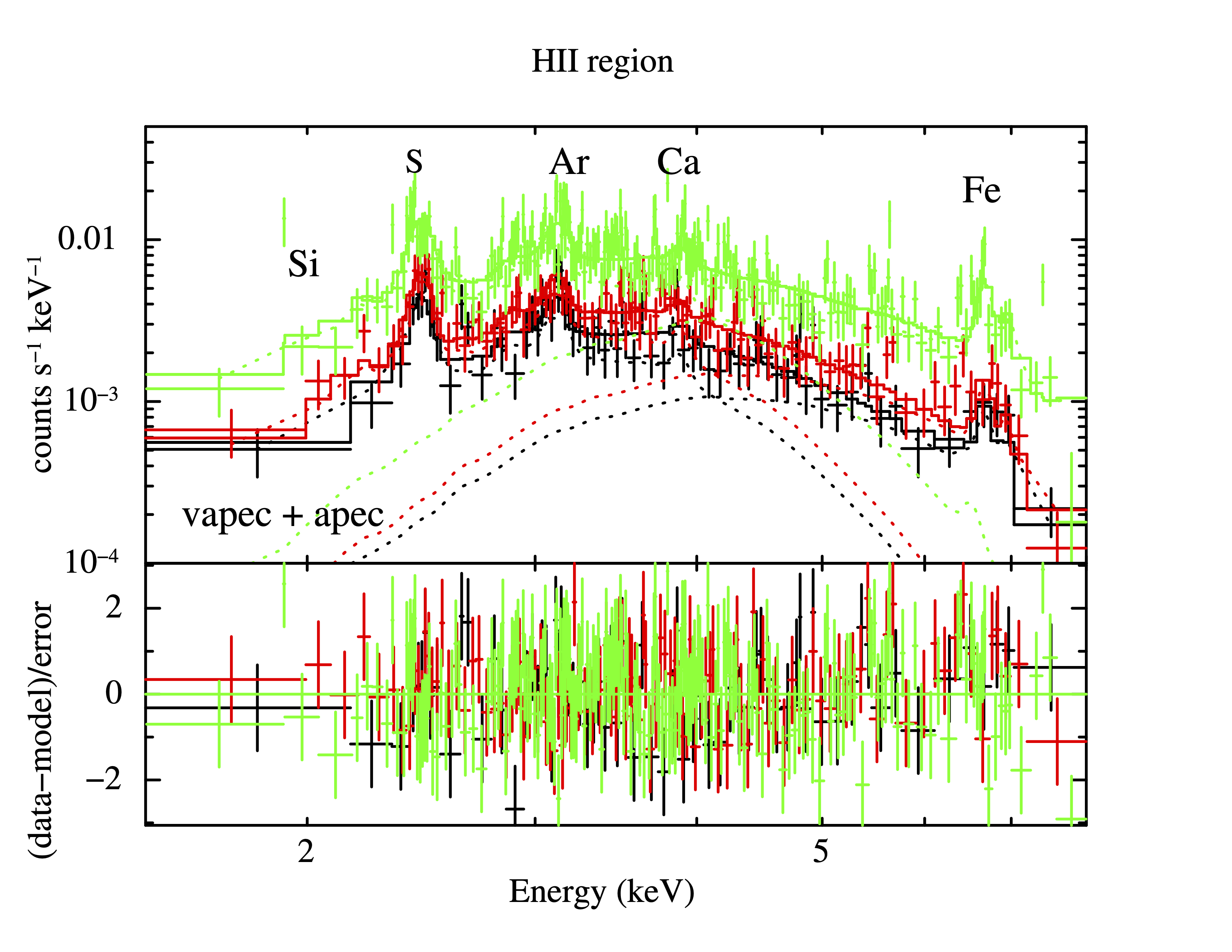}
\includegraphics[width=0.48\textwidth]{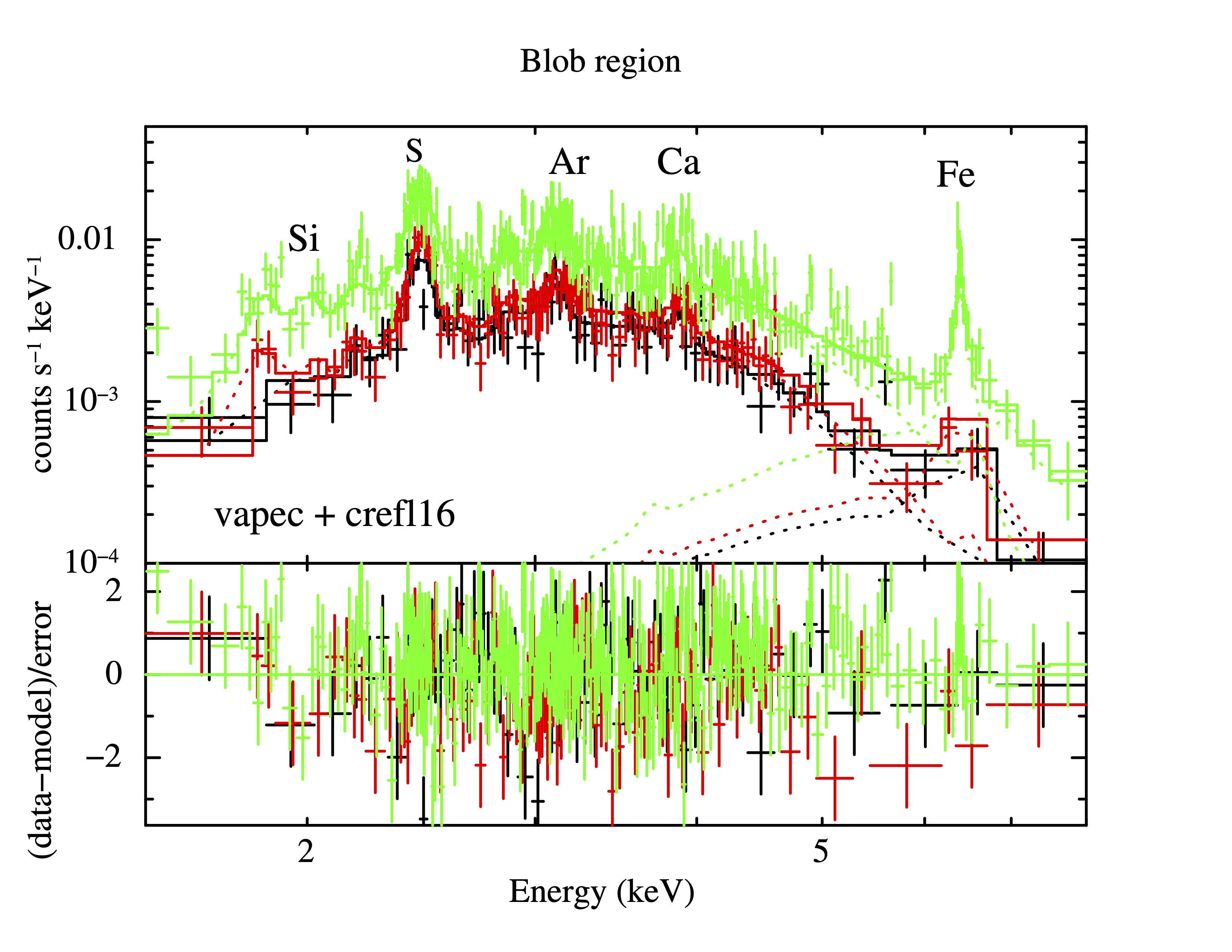}
\caption{{\it XMM-Newton} spectra of the Sgr C H\,\textsc{ii} region (left panels) and the Blob region (right panels). MOS1 in black, MOS2 in red and PN in green.\textit{Top Left:} H\,\textsc{ii} spectrum in 5--8 keV fitted using absorbed \texttt{vapec} model; \textit{Top Right:} Blob spectrum in 5--8 keV fitted with absorbed \texttt{vapec} model; \textit{Bottom Left:} Best-fit \texttt{tbabs*(vapec+apec)} model for HII region; \textit{Bottom Right:} Best-fit \texttt{tbabs*(vapec+crefl16)} model for Blob region; In bottom panels, the dotted lines show the individual components of the best-fit model.}
\label{fig:spectra}
\end{figure*}

\section{Spectral analysis} 
\label{sec:spectra}
We study the diffuse emission in the Sgr C complex through a spectroscopic analysis of the {\it XMM-Newton} data. 
To investigate the difference between the H\,\textsc{ii} region and the brighter Blob, we extracted spectra separately from these two circular regions.
The local background was estimated by averaging the spectra of three source-free regions surrounding the Sgr C complex (see the dotted green circles in Figure \ref{fig:multi_fig}) and subtracting that background from the source spectra, with proper weighting for area.

Spectral modeling was carried out using \texttt{XSPEC} \citep[v12.14,][] {Arnaud1996}. 
The spectra were binned to have at least one count per channel. 
Initially, we attempted to model the spectra as an absorbed thermal plasma in collisional ionization equilibrium using the \texttt{tbabs*vapec} model.
However, as shown in top-left panel of Figure \ref{fig:spectra}, a significant residual in the Fe K$\alpha$ band (6--8 keV) emerged in the spectral fitting result for the H\,\textsc{II} region.
We also tested the non-equilibrium (\texttt{vvnei}) and recombining plasma models (\texttt{vvrnei}) for the cool component, but saw no improvement in the fitting statistics. 
By adding an additional higher-temperature \texttt{apec} component, the fitting statistics were improved, and more importantly, the residuals in the Fe K$\alpha$ band become more evenly distributed.

When modelling the spectrum of the Blob with a simple absorbed thermal model, \texttt{tbabs*vapec}, we noticed an excess of 6.4 keV Fe K$\alpha$ line emission as shown in top-right panel of Figure \ref{fig:spectra}, which is highly likely to be contributed by reflection emission, as reported in \citet{Chuard2018} (See Appendix \ref{sec:appendix_reflection} for more details). 
We therefore added an additional reflection model, \texttt{crefl16} \citep[uniform Cloud REFLection model version of 2016;][]{Churazov2017}, to the absorbed thermal plasma model.
The model calculates the reflected spectrum of a uniform spherical cloud illuminated by a power-law spectrum with a photon index $\Gamma$, extending up to 1 MeV.

For spectral fitting, we set free the column density, plasma temperatures, abundances and the normalizations for each component.
We fit jointly the EMOS and EPN spectra, minimizing the C-statistic \citep{Cash1979}.
The protosolar abundance table from \citet{Lodders2009} has been applied for the abundances.
We show the spectra of both the H\,\textsc{ii} region and the Blob in Figure \ref{fig:spectra}, along with the best-fit models and text labels for prominent emission lines. 
Table \ref{tab:model} presents the best-fit parameters and their corresponding 1-$\sigma$ uncertainties.  
Both features -- the H\,\textsc{II} region and the Blob -- exhibit a cool plasma component with $kT \lesssim 1$ keV and similar abundances, suggesting a common origin.
Assuming a common origin for both features, we jointly fitted the spectra of the H\,\textsc{II} region and the Blob, tying the temperature ($kT_{\mathrm{c}}$) and metallicity of their cool plasma components. 
As presented in the right part of Table~\ref{tab:model}, the joint fit yields an acceptable statistical result with a consistent thermal plasma model characterized by $kT \sim 0.84 \pm 0.03$ keV and $Z \sim 1$–$2.5\,Z_{\odot}$. 

To further investigate the spatial variation of abundance in the H\,\textsc{II} region, we attempted to divide the circular region into four sectors (NE, NW, SE, SW). However, within the limited statistical uncertainties, no significant abundance variations are observed between the various sectors.


\begin{deluxetable*}{lcc|cc}[!thbp]
\tablecaption{Spectral fitting models and parameters\label{tab:model}}
\tablehead{
\colhead{Parameter} & \colhead{H\,\textsc{ii} region} & \colhead{Blob} & \colhead{H\,\textsc{ii} region$^{\dagger}$} & \colhead{Blob$^{\dagger}$}
}
\startdata
Model                  & \textit{tbabs*(vapec + apec)} & \textit{tbabs*(vapec + crefl16)} & \textit{tbabs*(vapec + apec)} & \textit{tbabs*(vapec + crefl16)} \\
Cstat / dof            & 2150.40 / 2159       &2315.93 / 2158    & \multicolumn{2}{c}{4475.90 / 4320}  \\
$N_{\rm H}$ ($10^{23}\,\rm cm^{-2}$) & 1.99 $\pm$ 0.12 & 1.77 $\pm$ 0.09 & 1.88 $\pm$ 0.08 & 1.84 $\pm$ 0.07  \\
$kT_{c}$ (keV)         & 0.77 $\pm$ 0.03       & 0.90 $\pm$ 0.04 & \multicolumn{2}{c}{$0.84\pm0.03$} \\
$kT_{h}$ (keV)         & 8.27 $\pm$ 0.78       & \nodata    & 8.11 $\pm$ 0.89  &   \nodata     \\
norm$_{c}$ ($\times 10^{-2}$)  & $1.63\pm0.20$   & $1.20\pm0.10$  & $1.14\pm0.2$  &   $1.83\pm0.26$ \\
norm$_{h}$ ($\times 10^{-2}$)  & $0.05\pm0.01$  & \nodata & $0.05\pm0.01$ & \nodata \\
norm$_{r}$ ($\times 10^{-2}$)  & \nodata  & $1.09\pm0.10$  & \nodata  & $1.17\pm0.09$  \\\hline
Si                     & $1.89\pm0.50$         & $2.71\pm0.80$   & \multicolumn{2}{c}{$2.41\pm0.62$}  \\
S                      & $1.22\pm0.20$           & $1.75\pm0.24$  & \multicolumn{2}{c}{$1.51\pm0.18$} \\
Ar                     & $1.40\pm0.31$           & $0.95\pm0.20$  & \multicolumn{2}{c}{$1.14\pm0.17$}  \\
Ca                     & $1.08\pm0.44$           & $1.13\pm0.26$  & \multicolumn{2}{c}{$1.24\pm0.23$}   \\
\enddata
\tablecomments{Temperatures are in keV; abundances are relative to solar. norm$_{h}$ and norm$_{r}$ represent the normalization of the hot component in the H\,\textsc{ii} region and the reflection emission in the blob region respectively. The quoted
errors are at 1$\sigma$ (68.3\%) confidence level.
$^{\dagger}$In the rightmost 2 columns, the spectra of the H\,\textsc{II} 
region and the Blob are jointly fitted, with the cool plasma temperature ($kT_{\rm c}$) and the elemental abundances in the \textit{vapec} component tied between the two regions.
}  
\end{deluxetable*}

\section{Discussion}
\label{sec:discussion}

It is not a rare case that we find an X-ray counterpart to the H\,\textsc{ii} region. 
Observations with \textit{Chandra}, \textit{XMM-Newton}, and \textit{Suzaku} have revealed diffuse X-ray emission associated with massive H\,\textsc{ii} regions such as the Carina Nebula \citep[e.g.,][]{Hamaguchi2007, Townsley2011} and 30\,Doradus \citep[e.g.,][]{Townsley2006}.
Stellar winds from O- and B-type stars, as well as colliding-wind binaries, can drive shocks that heat gas to X-ray emitting temperatures (10$^{6}$–10$^{7}$ K), while expanding wind-blown bubbles and superbubbles often exhibit diffuse, soft X-ray emission \citep[e.g.,][]{Townsley2003, Chu2008}.

\subsection{A SNR?}
\label{sec:diss-snr}

As demonstrated in \citet{Riquelme2025}, the velocity-integrated [C\,\textsc{ii}] map reveals an expanding shell surrounding the H\,\textsc{ii} region. 
The massive shell expands at an unexpectedly high velocity of $\sim$23 km s$^{-1}$, exceeding what stellar feedback alone can plausibly explain and suggesting alternative drivers such as a SNR.

A supernova could be a candidate for producing the X-rays within the H\,\textsc{ii} region, and given the presence of massive stars there \citep{Nogueras-Lara2024}, the most likely type of supernova to occur there would be a core-collapse supernova.  However, if a supernova had occurred within the H\,\textsc{ii} region, we could expect it to have enhanced the metallicity, at least of the X-ray emitting plasma.

Previous studies have indicated that there is a clear metallicity gradient within the Galaxy and that the metal abundance of gas in the CMZ is on the order of twice solar \citep[e.g.,][]{Simpson+95, Giveon+02b, Martin-Hernandez+02, Rudolph+06}. Indeed, studies of the innermost few parsecs of the Galactic center exhibit an overall metal enrichment roughly on the order of twice solar, as indicated by both X-ray studies \citep{Hua2023, Hua2025} and JWST measurements \citep{Vermot2025}, where winds from Wolf–Rayet stars dominate the chemical enrichment.  The abundances of Si, S, and Ar inferred from our X-ray study of both the Sgr C H\,\textsc{ii} region and the Blob appear to be roughly consistent with the expected super-solar metallicity of the interstellar medium in the CMZ, although they do not exceed it.

Moreover, strong absorption in the soft X-ray band ($E \lesssim 1.5$ keV) prevents robust constraints on O and Mg, which are among the most sensitive tracers of metal-rich supernova ejecta. Taken together, these results do not provide compelling evidence for a chemically young, ejecta-dominated supernova remnant within the H \textsc{ii} region.

Nevertheless, the absence of a clear ejecta signature does not preclude the presence of an evolved SNR whose X-ray emission is dominated by shock-heated ambient gas rather than metal-rich ejecta. In such a scenario, we can estimate the shock velocity from the observed plasma temperature using
\begin{equation}
v_s=\sqrt{\frac{16 k T}{3 \mu m_p}}=\left(923 \mathrm{~km} \mathrm{~s}^{-1}\right)\left(\frac{k T}{1 \mathrm{keV}}\right)^{\frac{1}{2}},
\label{eq:shock_velocity}
\end{equation}
where $\mu$ = 0.6 is the mean atomic weight for a fully ionized plasma.
Since the hotter component shows no evidence of metal enhancement, and since supernova remnants almost always display temperatures well below 8 keV, 
this component is unlikely to be physically associated with the SNR (see Section~\ref{sec:diss-hot} for a more detailed discussion). We therefore consider only the cooler component in the following analysis. Substituting kT = 0.77 keV into equation \ref{eq:shock_velocity}, we obtained a shock velocity of $v_s \approx 800\,\mathrm{km\,s^{-1}}$ for the cooler component.

Assuming that gas fills a sphere within the spectral extraction region ($R=1\farcm{5}$), we derive a gas density of $n_{\rm H}=1.4\pm 0.1~\rm cm^{-3}$ and a post-shock electron density of $n_{\rm e}= 1.2n_{\rm H} = 1.7\pm 0.1~\rm cm^{-3}$.
The corresponding gas mass is $M_X=8.7\pm 0.5~M_\odot$ \footnote{If the X-ray emission arises from a thin shell with a thickness 1/12 of the SNR radius, the density and mass would be revised to $n_{\rm H}=2.9\pm 0.2~\rm cm^{-3}$ and  $M_X=4.2\pm 0.3~M_\odot$, respectively. Nevertheless, the X-ray image does not reveal a shell-like morphology for this source.}.  The X-ray-emitting gas mass is comparable to the typical ejecta mass from a core-collapse supernova.
As noted above, the observed abundances in the X-ray emitting gas associated with the H\,\textsc{ii} region are only mildly elevated above solar values, and are comparable to the metallicity expected for the ambient gas in the CMZ, suggesting that, if a supernova is responsible for the X-ray emission and for the expansion of the H\,\textsc{ii} region, then a substantial amount of ambient gas has mixed with the SN ejecta, and only a small proportion of the ejecta has been observed. This would imply the existence of a gas component even cooler than 0.77 keV that contains most of the ejecta. Due to the strong absorption, the X-ray spectra do not allow us to determine whether this colder component exists or not.

Assuming that the SNR is in the Sedov-Taylor phase, we roughly estimate the SNR age to be $t=0.4 R/V_b\sim   1700(V_b/800~\rm km~s^{-1})^{-1}$~yr, where $V_b$ is the blast wave shock velocity. Considering an undetected cooler gas component, $V_b$  could be slower than the shock velocity we infer, i.e., $V_b \lesssim 800~\rm km~s^{-1}$. Therefore, the 1700~yr should be regarded as a lower limit to the SNR age.
The inferred age is broadly consistent with the population of young SNRs in the Galactic Center, such as Sgr A East \citep{Zhou2021, Zhang2023}, G359.0$-$0.9 \citep{Bamba2000}, and G0.9+0.1 \citep{Venter2007}, whose ages lie in the $10^3$–$10^4$ yr range. However, this lower limit is much smaller than the upper limit on the expansion age of the C$^+$-emitting shell around the HII region \citep{Riquelme2025}, which could have been set in motion by one or more past supernovae.

Although the current evidence is consistent with a SNR interpretation, the possibility of a non-SNR origin cannot be ruled out. As discussed above, the
metal enhancement implied by the X-ray spectrum is only marginal, particularly given the locally super-solar abundance environment, so there is no evidence that a supernova has enhanced the local metallicity. This contrasts with G359.1-0.5, a well-established core-collapse SNR in the Galactic Center, which exhibits strong metal enrichment—over ten times the solar abundance for both S and Si \citep{Ohnishi2011}.
Additionally, a radio shell is usually observed surrounding the central diffuse X-ray emission in known SNRs; however, no such shell-like structure is evident in the current radio morphology.

As an alternative scenario, the diffuse X-ray emission might be attributed to the massive stars within the H\textsc{ii} region. However, for comparison, the massive, young Arches and Quintuplet clusters have absorption-corrected X-ray luminosities of $L_{\rm 2–8~keV} \sim 3 \times 10^{33}~\rm erg~s^{-1}$ \citep{Wang2006}. In contrast, based on the surface brightness measured in Section~\ref{sec:image}, the Sgr C H\textsc{ii} region exhibits a substantially higher luminosity of $L_{\rm 2–8~keV} \sim 4.2 \times 10^{34}~\rm erg~s^{-1}$, exceeding that of both clusters by more than an order of magnitude.

\begin{figure}[!tbh]
\centering
\includegraphics[width=0.45\textwidth]{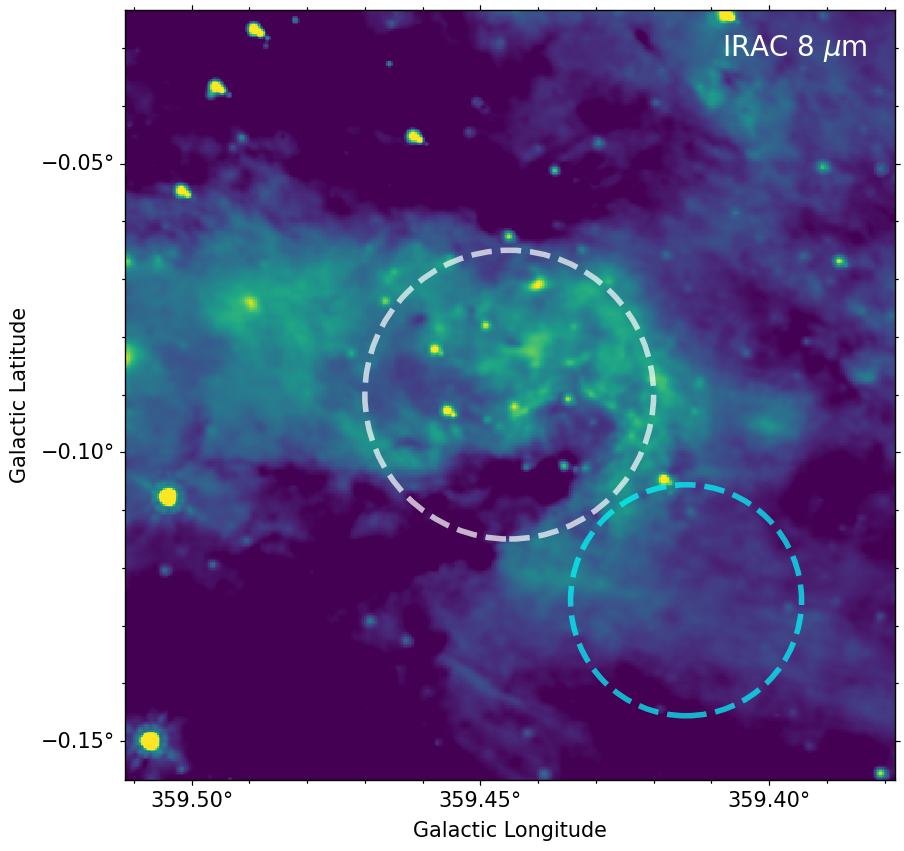}
\includegraphics[width=0.45\textwidth]{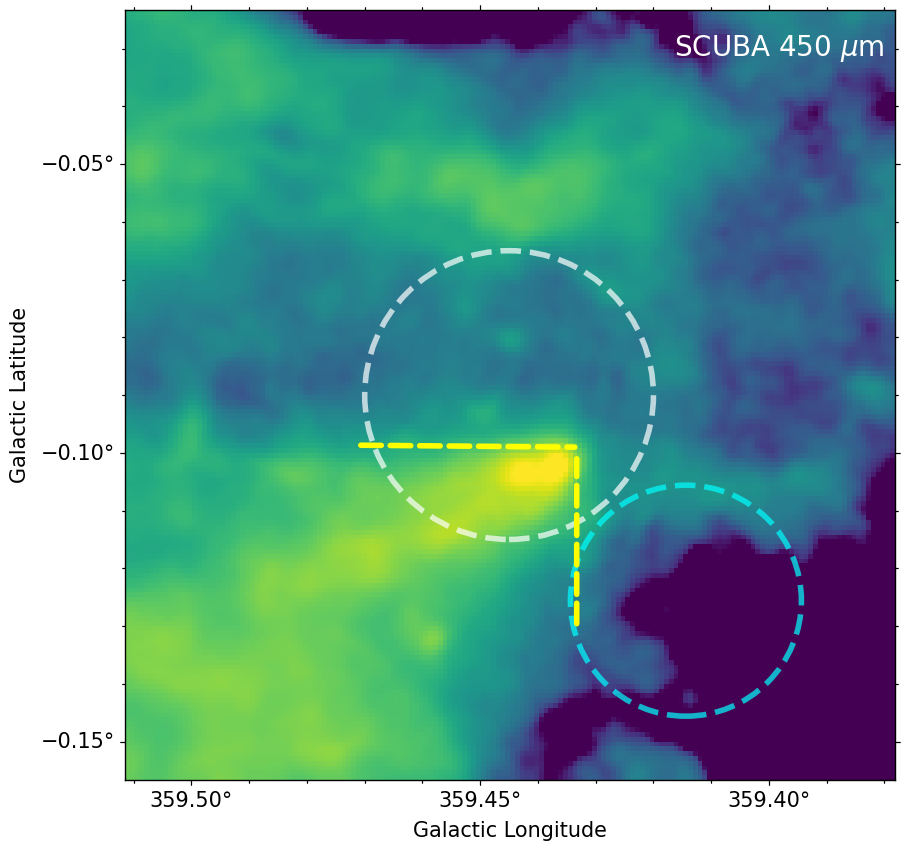}
\caption{Dust distribution in the SgrC complex. 
\texttt{Upper:} {\it Spitzer}/IRAC 8 $\mu$m view; \texttt{Lower:} {\it JCMT}/SCUBA 450 $\mu$m view. The white circle marks the H \textsc{ii} region, while the cyan circle shows the Blob. The dashed yellow lines mark the edges observed in the X-ray images shown in Figure \ref{fig:xray_fig}.
}
\label{fig:infrared}
\end{figure}

\subsection{The hot component in the H\,\textsc{ii} region}
\label{sec:diss-hot}
It is noteworthy that we find evidence for a hard thermal component, $\sim$8 keV, in the H\,\textsc{ii} region. 
To first assess whether this hot component arises from the Galactic Ridge X-ray Emission (GRXE), we compared spectra extracted from local backgrounds near the Galactic plane with those at higher latitudes, where the GRXE contribution decreases with increasing distance from the plane.
The $\sim$8 keV excess persists even after adopting the revised local background, indicating that its origin is unlikely to be associated with the GRXE.

As discussed in Section~\ref{sec:diss-snr}, the hot component could be associated with the SNR. In the following, we discuss two other possible origins.
This high-energy component might come from strong shocks where the strong winds of the WCL Wolf-Rayet stars (WRs) impact the dense shell surrounding the H \textsc{ii} region.
A K-band spectroscopic survey of over 500 highly reddened point sources in the CMZ carried out by \citet{Geballe2019} revealed two WRs (17443734-2927557 and 17444083-2926550) in or near the H\,\textsc{ii} region of the Sgr C complex.
By fitting the broadening of the He\,\textsc{i} emission lines, we obtained a rough estimate of the wind speed for both stars on the order of $\sim$ 2000 km s$^{-1}$ (see the spectra presented in Figure \ref{fig:WRs} of Appendix \ref{sec:appendix_WR} ). 
Assuming that the bulk kinetic energy of the upstream flow is fully converted into thermal energy of the post-shock gas, we estimate the gas temperature using Equation~\ref{eq:shock_velocity}.
For a pre-shock velocity of 2000~km~s$^{-1}$, this yields a post-shock temperature of $kT \simeq 4.7$~keV, representing an upper limit on the X-ray temperature attainable in this scenario. This mechanism therefore seems to fall short of being a viable source of the 8 keV component of the spectrum.

Alternatively, this hot component could arise from colliding winds of massive stars in the H\,\textsc{ii} region.
Similarly to W49A, an ultracompact H\,\textsc{ii} region containing a dozen nascent early-type stars, {\it Chandra} has detected hard X-ray plasma $\sim$ 7 keV \citep{Tsujimoto2006}.
The authors interpreted the observed hard X-ray emission from W49A in terms of an ensemble of point sources -- O-type stars -- of shocked interacting winds, and a wind-blown bubble interacting with ambient cold matter. 
A similar scenario could be applicable to the Sgr C H\,\textsc{ii} region, if such stars are found to be present.
A recent investigation by \citet[][and personal communication]{Nogueras-Lara2024} shows that the immediate environment of the Sgr C H\,\textsc{ii} region hosts $\sim10^{5}$ M$_{\odot}$ of young stars with minimum ages of $\sim20$ Myr.
The winds from the most massive of these stars could interact with the strong stellar winds of the WR stars and produce the hot X-ray emission.

\subsection{The offset X-ray ``Blob"}
\label{sec:diss-blob}

In X-ray morphology, the bright Blob appears physically connected to the nearby H\,\textsc{ii} region. 
The comparable column densities and the presence of a similarly cool thermal component in both regions suggest that they may share a common origin. 
The consistently high foreground column densities also exclude the possibility that either component arises from a foreground source. 
The relatively low hardness ratio of the Blob further hints that it could represent material outflowing from the H\,\textsc{ii} region. However, this interpretation is challenged by the fact that the Blob is significantly brighter in X-rays than the H\,\textsc{ii} region itself, which is difficult to reconcile with a simple outflow scenario.

To further investigate why the Blob exhibits a higher surface brightness, we examine near- and far-infrared images that trace both warm and cold dust. The Spitzer/IRAC 8 µm map \citep{Stolovy2006, Arendt2008} shows warm dust concentrated around the H \textsc{ii} region, with much fainter emission toward the Blob (Figure \ref{fig:infrared}). Similarly, the JCMT/SCUBA 450 µm image \citep{Pierce-price2000} reveals strong far-infrared emission from cold dust toward the H \textsc{ii} region, whereas a clear deficit of cold dust is seen along the line of sight to the Blob. 
For a more detailed discussion of the infrared morphology of Sgr C—including prominent features visible in Figure \ref{fig:infrared}, such as the G359.44–0.10 EGO—see \citet{Zhao2025}.
The dust distribution provides a natural explanation for why the Blob appears brighter than the H \,\textsc{ii} region: foreground dust absorption suppresses the X-ray emission from the H \,\textsc{ii} region, and reshapes the overall X-ray morphology of the Sgr C complex.
We also note that the boundaries of the X-ray–dark regions closely follow the dense cold-dust structures traced by SCUBA (see yellow lines in both Figure \ref{fig:xray_fig} and Figure \ref{fig:infrared}), suggesting that intrinsic X-ray emission may be emitted by the southeastern part of the Sgr C H\,\textsc{ii} region and its immediate surroundings, but is obscured by intervening dust.

An alternative possibility is that the Blob is a separate source that is merely projected adjacent to the H\,\textsc{ii} region along the line of sight. The Blob is infrared-dark, as reported by \citet{Zhao2025}, who reported a strong spatial anti-correlation between this southwest X-ray peak and far-IR emission. It is also radio-dark, as indicated by the deep red region shown in the left panel of Figure~\ref{fig:multi_fig}. Since no convincing counterpart has been detected at any other wavelength, the physical origin of the X-ray emission from the Blob remains unclear, but the hypothesis that the emitting plasma in the Blob has been extruded from the H\,\textsc{ii} region remains as a possibility.

As the successor to ASTRO-H ({\it Hitomi}), the XRISM satellite provides unprecedented access to high-resolution X-ray spectroscopy. Its onboard micro-calorimeter ({\it Resolve}) achieves an energy resolution of $\leq$5 eV across the 0.3–12 keV band, enabling precise measurements of gas velocities and turbulent motions. A new XRISM pointing toward the Sgr~C complex was obtained earlier this year, and the forthcoming spectral diagnostics may reveal crucial kinematic information—such as relative bulk velocities between the H,\textsc{ii} region and the Blob—which could help distinguish between a physical association and a line-of-sight superposition.

\subsection{The extended Channel}
\label{sec:channel}
Previously discovered and named as the ``Chimney" by \citet{Tsuru2009}, this diffuse feature actually extends further in latitude than they had reported, being limited by the {\it Suzaku} field of view.
\citet{Tsuru2009} suggested that the Channel is physically connected with the unresolved main structure comprised of both the H\,\textsc{ii} region and the Blob, based on their measured spectral properties.

With our {\it XMM-Newton} image as shown in Figure \ref{fig:chimney}, we find that the Channel extends 26 pc northward from the H\,\textsc{ii} region, and has a maximum width of 12 pc.
The feature appears very soft, exhibiting an observed surface brightness of SB$_{1.5\text{--}4\,\mathrm{keV}} = 3.0 \times 10^{-14}\ \mathrm{erg\ s^{-1}\ cm^{-2}\ arcmin^{-2}}$, which is only 15\% higher than the local background.
However, at higher energies ($>$ 4 keV), we detect no significant flux excess in the Channel. 
The limited photon statistics of the Channel prevent a reliable determination of the Channel’s spectral properties from the {\it XMM-Newton} data.

The fact that the Channel is within a few degrees of being perpendicular to the Galactic plane raises the possibility that the X-ray-emitting plasma in the Channel is being collimated by the relatively strong vertical magnetic field present in the Central Molecular Zone, evidenced by the forest of predominantly vertical filaments of nonthermal radio emission, or NTFs \citep{YZ+04, Morris06c, Heywood+22}.  These NTFs delineate and illuminate the magnetic field in the intercloud medium where relativistic electrons are produced \citep{Morris15}. It is therefore noteworthy that one of the most prominent vertical NTFs is superimposed upon the Channel (compare Figures \ref{fig:chimney} and \ref{fig:multi_fig}), and has an orientation that differs by only about 10 degrees from that of the Channel.  That Sgr C NTF is therefore a candidate for being a synchrotron-illuminated part of the magnetic field that confines the Channel.  The neighboring and much larger-scale X-ray Chimney has also been considered to be confined by the vertical magnetic field \citep{Ponti2019, ZhangMF+21}.

\section{Conclusion}
\label{sec:conclusion}
In this work, we have analyzed the available {\it XMM-Newton} and {\it Chandra} observations of the Sgr C complex. Compared to previous Suzaku studies, our analysis reveals the detailed X-ray morphology of Sgr C with improved spatial resolution, showing that it consists of two thermally emitting regions: the H \textsc{ii} region and a bright Blob. We find that:
\begin{itemize}
    \item The H \textsc{II} region shows an abundance pattern broadly consistent with the elevated metallicity expected in the CMZ. When considered alongside the expanding [C \textsc{II}] shell observed in the far-infrared and possible mixing between the ambient gas and supernova ejecta, the evidence is consistent with a SNR origin for the X-ray emission. 
    \item  We measure a $\sim$0.8 keV post-shock plasma with an electron density of $1.7 \pm 0.1\ \mathrm{cm^{-3}}$ for the H \textsc{II} region. Under the SNR interpretation, assuming the remnant is in the Sedov–Taylor phase, we estimate an age of $\sim$ 1.7 kyr for the SNR. These values are broadly consistent with a core-collapse supernova origin in the Galactic Center.
    \item The bright Blob exhibits higher surface brightness than the H \textsc{II} region.  Although this is difficult to reconcile with a scenario in which the emitting plasma in the Blob is outflowing from the HII region, a joint fit of the spectra of the two sources indicates very similar temperatures and foreground column densities, so a common origin for these sources remains plausible. 
   \item Compared to previous studies, our {\it XMM-Newton} analysis provides a more complete view of the elongated vertical Channel, which extends $\sim$26 pc northward from the H\,\textsc{ii} region and reaches a maximum width of $\sim$12 pc.
\end{itemize}

\begin{acknowledgments}
We thank the anonymous referee for constructive suggestions that improved this paper.
We thank Tom Geballe for providing the spectra of the WR stars and Francisco Nogueras-Lara for useful information on the young star population in the vicinity of Sgr C. 
Z.Z. and M.R.M. acknowledge financial support from NASA ADAP grant 80NSSC24K0639 to UCLA.
G.P. acknowledges financial support from the Framework perl’Attrazione e il Rafforzamento delle Eccellenze (FARE) per la ricerca in Italia (R20L5S39T9). 
G.P. also acknowledges financial support from the European Research Council (ERC) under the European Union’s Horizon 2020 research and innovation program HotMilk (grant agreement No.865637). 
P.Z. thanks the support by NSFC 12273010.
This research employs a list of Chandra datasets, obtained by the Chandra X-ray Observatory, contained in~\dataset[doi:10.25574/cdc.554]{https://doi.org/10.25574/cdc.554}.
\end{acknowledgments}

\facilities{XMM, CXO, MeerKAT, UKIRT/UIST, SOFIA/upGREAT, Spitzer/IRAC, JCMT/SCUBA}

\software{astropy~\citep{2013A&A...558A..33A,2018AJ....156..123A,2022ApJ...935..167A},  APLpy~\citep{Robitaille2012}, DS 9~\citep{Joye2003}, CIAO~\citep{Fruscione2006}, SAS
          }

\bibliography{sgrc}{}
\bibliographystyle{aasjournalv7}

\appendix
\restartappendixnumbering
\section{Reflection emission}
\label{sec:appendix_reflection}
To investigate the reflection component in our regions of interest, we examine the 6.32–6.48 keV flux map overlaid with the Fe K$\alpha$–emitting clumps identified by \citet{Chuard2018}, namely Sgr C1, Sgr C2, and Sgr C4. 
Because the reflection component is time-variable, we have slightly adjusted the centers and sizes of these clumps to more accurately match our {\it XMM} mosaic combined from observations spanning 2000 to 2020.
A significant portion of the bright Blob overlaps with Sgr C4, whereas the H \textsc{ii} region shows only slight spatial coincidence with Sgr C1. 
This spatial configuration helps explain why a residual reflection component is present in the spectrum of the Blob but not in that of the H \textsc{ii} region.

\begin{figure}[!tbh]
\centering
\includegraphics[width=0.49\textwidth]{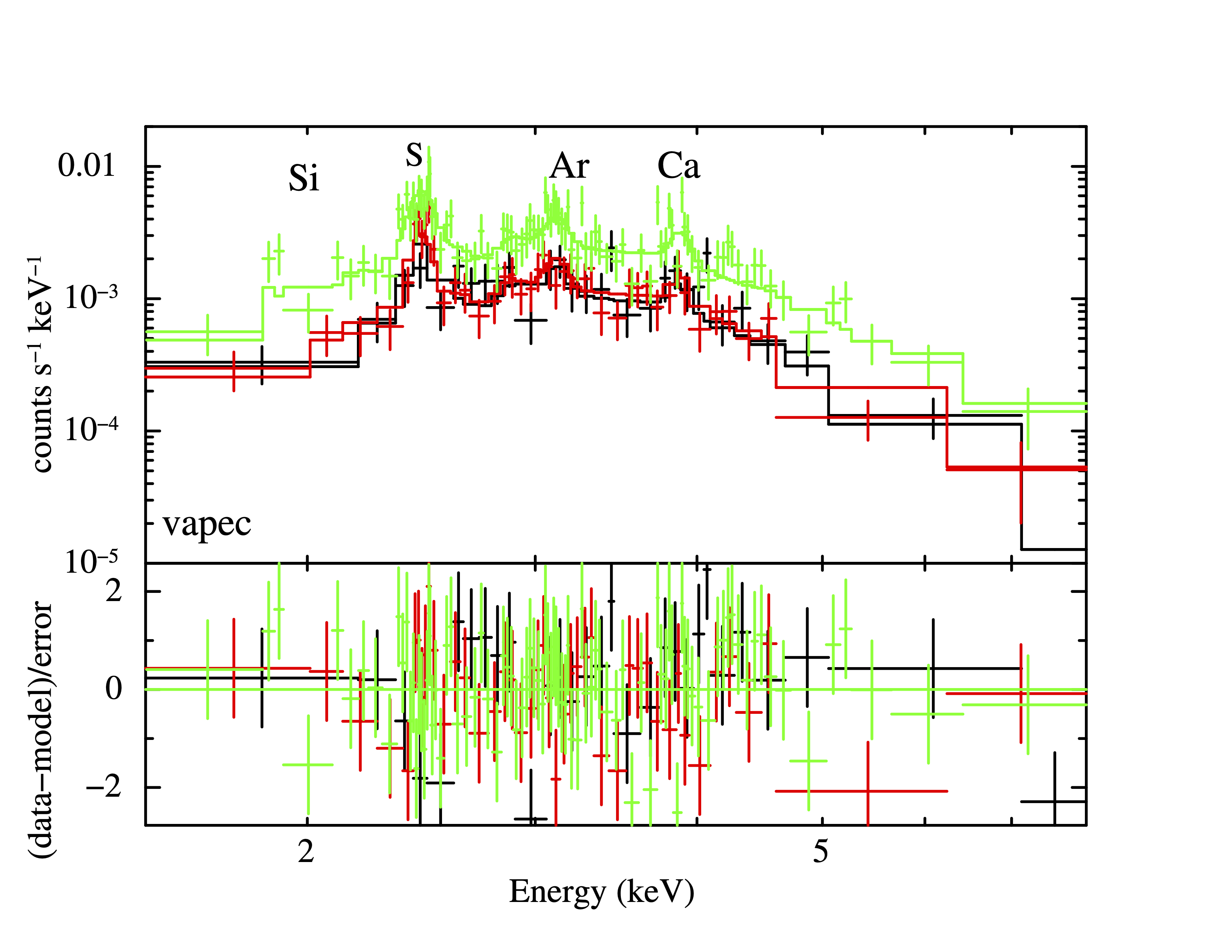}
\caption{Spectrum extracted from the portion of the Blob that does not overlap with Sgr C4.}
\label{fig:blob_out_of_C4}
\end{figure}

To check this further, we extracted the spectrum from the Blob region excluding the area overlapping with Sgr C4. The spectrum was fitted with the model \textit{tbabs*(vapec+crefl16)}. As shown in Figure \ref{fig:blob_out_of_C4}, the best-fit normalization of the reflection component is $(1.38 \pm 0.50) \times 10^{-3}$, which is not statistically significant at the $3\sigma$ level. This result, plus the known  spatial and temporal variability of the reflection emission, supports the conclusion that the 6.4 keV Fe K$\alpha$ emission originates from reflection by the molecular cloud.

\begin{figure}[!tbh]
\centering
\includegraphics[width=0.48\textwidth]{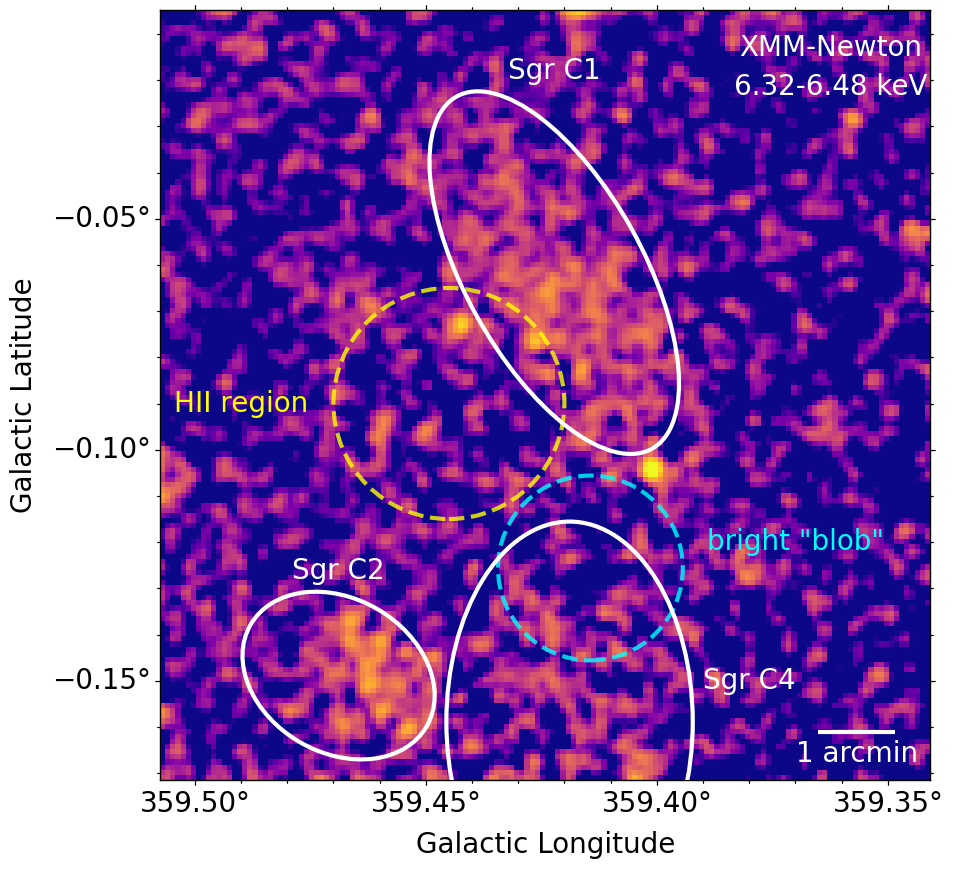}
\caption{Fe 6.4 keV flux map of the Sgr C complex. The dashed circles mark the two regions of interest: the H \textsc{ii} region (yellow) and the Blob (cyan). }
\label{fig:reflection}
\end{figure}

\restartappendixnumbering

\section{Wind speed of the WR stars}
\label{sec:appendix_WR}
\begin{figure}[!tbh]
\centering
\includegraphics[width=0.48\textwidth]{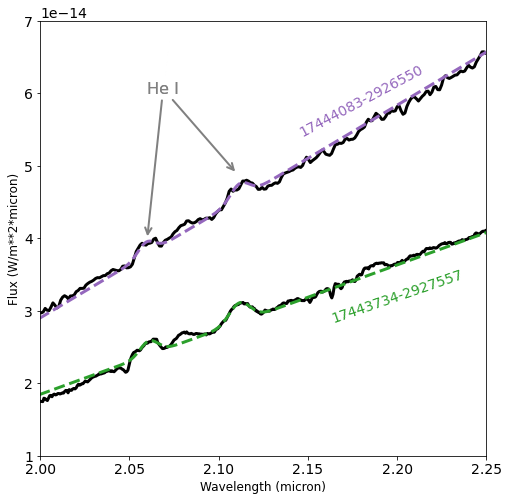}
\caption{K-band spectra of two Wolf-Rayet stars in the H\,\textsc{ii} region of SgrC, taken with the United Kingdom Infrared Telescope \citep{Geballe2019}. The dashed lines show best-fit models composed of a power-law and two Gaussian lines. }
\label{fig:WRs}
\end{figure}

In Figure~\ref{fig:WRs}, we present the K-band spectra of the two WR stars (17443734–2927557 and 17444083–2926550) located within or at the edge of the H\,\textsc{ii} region in Sgr C. To obtain an approximate estimate of their wind velocities, we fitted the spectra with a power-law continuum and two Gaussian components to model the He I emission line. 
From the best-fit FWHM of the Gaussian profiles, we derive wind velocities on the order of $\sim$2000 km s$^{-1}$ for both stars. 

\end{document}